\documentclass[apj]{emulateapj}
\usepackage{amsmath}
\usepackage{tmsfnts}

\def\kmsmpc     {km$\;$s$^{-1}\,$Mpc$^{-1}$}
\def\wmap       {\emph{WMAP}}
\def\chandra    {\emph{Chandra}}

\begin{document}

\submitted{Submitted to ApJ Letters, 2007 May 23}

\lefthead{HELIUM ABUNDANCE IN GALAXY CLUSTERS}
\righthead{MARKEVITCH}

\title{HELIUM ABUNDANCE IN GALAXY CLUSTERS AND SUNYAEV-ZELDOVICH EFFECT}

\author{M.~Markevitch\altaffilmark{1}}

\affil{Harvard-Smithsonian Center for Astrophysics, 60 Garden St.,
Cambridge, MA 02138; maxim@head.cfa.harvard.edu}

\altaffiltext{1}{Also Space Research Institute, Moscow, Russia}

\setcounter{footnote}{1}

\begin{abstract}

It has long been suggested that helium nuclei in the intracluster plasma can
sediment in the cluster gravitational potential well. Some theoretical
estimates for the cores of relaxed clusters predict an excess of helium
abundance by up to a factor of a few over its primordial value.  The
intracluster helium abundance cannot be measured directly. This presents a
significant source of uncertainty for cosmological tests based on the X-ray
derived cluster quantities, such as the gas mass, total mass, and gas mass
fraction, all of which depend on the assumed helium abundance.  We point out
that cluster distances derived by combining the Sunyaev-Zeldovich (SZ) and
X-ray data also depend on the helium abundance. This dependence can be used
to {\em measure}\/ the abundance, provided the distance is known
independently.  For example, if one adopts the \wmap\ $H_0$ value, then the
recent $H_0$ measurement by Bonamente and collaborators, derived from SZ
data on 38 clusters assuming a primordial helium abundance, corresponds to
an abundance excess by a factor of $1.9\pm0.8$ within $r\sim 1$ Mpc (using
only their statistical errors).  This shows that interesting accuracy is
within reach. We also briefly discuss how the SZ and X-ray cluster data can
be combined to resolve the helium abundance dependence for the $d_a(z)$
cosmological test.

\end{abstract}


\section{INTRODUCTION}
\label{sec:intro}

The majority of helium in the universe, predominantly in the form of $^4$He,
was produced during the Big Bang. For the \wmap\ value of $\Omega_b
h^2=0.0223\pm 0.0007$ (Spergel et al.\ 2007), the standard hot big bang
nucleosynthesis model predicts a primordial fraction of helium in the total
baryonic mass density of $Y=0.2482\pm 0.0007$ (Walker et al.\ 1991; Kneller
\& Steigman 2004).  Recent spectral measurements in metal-poor extragalactic
H{\small II} regions give a value within 1\% of this theoretical prediction,
with similarly small uncertainties (Izotov et al.\ 2007; Peimbert et al.\ 
2007).  Thus, the primordial helium abundance appears to be known quite
accurately.

Helium abundance in the hot intracluster medium (ICM) may differ
significantly from the primordial one.  First, additional helium comes from
the stars.  The ratio of star mass to ICM mass is higher in the cluster
centers, so stellar enrichment will be stronger there.  However, the mass of
the primordial helium in the ICM is comparable to the total stellar mass in
a cluster, so helium enrichment by stars should not be significant (unlike
stellar contribution of heavier elements, which are present in the ICM in
trace amounts).  A much greater increase of helium abundance in the central
regions of clusters may be caused by sedimentation of heavy nuclei of the
ICM in the cluster gravitational potential (Fabian \& Pringle 1977; Rephaeli
1978; Abramopoulos, Chanan, \& Ku 1981; Gilfanov \& Sunyaev 1984; Qin \& Wu
2000; Chuzhoy \& Nusser 2003; Chuzhoy \& Loeb 2004; Ettori \& Fabian 2006).
The consensus of the recent works is that if sedimentation is not
suppressed, then in a hot cluster undisturbed for several gigayears, the
relative helium abundance can increase by a factor of 2 or more within
$r<0.2-0.3 r_{\rm 200}$, and even more at smaller radii (Chuzhoy \& Loeb
2004).  However, sedimentation can be inhibited by several mechanisms,
including tangled magnetic fields (which should also suppress diffusion and
thermal conduction in the ICM, as seems to be observed, e.g., Ettori \&
Fabian 2000; Vikhlinin et al.\ 2001; Markevitch et al.\ 2003), gas mixing by
cluster mergers and turbulence, and the formation of a cluster cool core
(Ettori \& Fabian 2006), because the diffusion rate is a strong function of
the temperature.

While mergers and turbulence should inhibit any contemporary sedimentation,
they are unlikely to permanently erase a large-scale abundance gradient
already present by the time of the disturbance.  The reason is the ICM in
relaxed clusters is stratified, with low-entropy gas at the bottom of the
gravitational well and higher-entropy gas in the outskirts. Such a stable
gas distribution should restore itself, and any radial abundance gradient
with it, shortly after a disturbance, provided that small-scale ICM mixing
during a merger is inefficient.  Indeed, we do observe radially declining
{\em iron}\/ abundance profiles in all relaxed clusters with sufficiently
detailed X-ray data, with the decline traced from the cluster cores to at
least $r\sim 0.5 r_{200}$ (e.g., Fukazawa et al.\ 1994; Tamura et al.\ 2004;
Vikhlinin et al.\ 2005).  While the iron abundance gradients are probably
caused by enrichment rather than sedimentation, they should be old enough to
have survived a merger or two, suggesting that it is difficult to erase an
abundance gradient. Thus, given the uncertainties in the processes that
inhibit sedimentation in the ICM, it is unclear how significant it would be
in a typical cluster.

As pointed out by Qin \& Wu (2000) and in later works (and summarized in
\S\ref{sec:fgas} below), helium abundance affects cluster quantities derived
from X-ray observations, such as the cluster total mass, gas mass, and gas
mass fraction (see also Belmont et al.\ 2005 for an application to the hot
gas in the center of our Galaxy).  The assumed helium abundance also affects
abundances of heavier elements derived from their X-ray emission lines
(e.g., Drake 1998; Ettori \& Fabian 2006).  Most of the current X-ray
cluster analyses are restricted to bright central regions --- precisely
those regions that may be affected by sedimentation.  Unfortunately, helium
in the ICM is fully ionized and not directly observable by spectroscopic
means. For this reason, its abundance is unknown and has to be adopted from
unrelated measurements, e.g., helioseismology (for a review see, e.g.,
Lodders 2003).  The widely used X-ray spectral fitting package {\small
  XSPEC} offers a choice of abundance models with the number density of
helium relative to hydrogen,
\begin{equation}
x\equiv \frac{n_{\rm He}}{n_p},
\end{equation}
spanning a range between 0.0792 (Lodders 2003) to 0.0977 (Anders \& Grevesse
1989 and some others).  For comparison, if one takes the abundances of
heavier elements to be $0.3-0.5$ solar (as in clusters), the CMB-based
primordial helium abundance (Spergel et al.\ 2007) corresponds to $x=0.083$.

The unknown cluster helium abundance is a source of uncertainty for X-ray
cluster-based cosmology studies. In this Letter, I propose a way to measure
it.

\section{X-RAY DERIVED QUANTITIES}

\subsection{Gas mass fraction}
\label{sec:fgas}

As mentioned above, a helium abundance in the ICM is implicitly assumed in
most X-ray derived cluster quantities, such as gas mass, total mass from the
hydrostatic equilibrium equation, and their ratio $f_{\rm gas}$.  All of
these are being used for cosmological tests (see, e.g., Vikhlinin et al.\ 
2003, Henry 1997, and Allen et al.\ 2004 for the three quantities,
respectively), with projects underway to use them for ``precision
cosmology''.  Below, their dependences on the assumed helium abundance are
written in a form relevant for the X-ray analysis (that is, fixing the X-ray
and other observables).  For simplicity, I assume a uniform abundance over
the region involved.  Current cosmological tests use relatively small
central cluster regions ($r<r_{500}$ or even $r<r_{2500}$), for which the
effects of sedimentation and enrichment can be significant.

In a fully ionized intracluster plasma, the number of electrons per proton
is
\begin{equation}
\frac{n_e}{n_p} =1+2x+x_{eh} \approx 1+2x,
\end{equation}
where $x_{eh}$ represents electrons from elements heavier than helium;
$x_{eh}\approx 0.005$ for the intracluster chemical abundances ($0.3-0.5$
solar), and we will neglect it. The mean molecular weight of the ICM is
\begin{equation}
\mu = \frac{1+4x+m_h}{2+3x+x_{eh}+x_{h}} \approx \frac{1+4x}{2+3x},
\label{eq:mu}
\end{equation}
where $m_h$ and $x_h$ are the mass and number density contribution from
elements heavier than helium. For clusters, $m_h\approx 0.01$ and can be
neglected; $x_{h}\ll x_{eh}$ and is certainly negligible as well.  Thus, for
clarity, we will ignore heavy elements below.

The total mass of a cluster within a certain radius, derived under the
assumption of hydrostatic equilibrium (e.g., Sarazin 1988) and (for our
illustrative purposes) isothermality, is
\begin{equation}
M_{\rm tot} \propto \frac{T_e}{\mu} \frac{d \log \rho_{\rm gas}}{d \log r}
            \propto \frac{1}{\mu} = \frac{2+3x}{1+4x}.
\label{eq:mtot}
\end{equation}
Here we used the fact that the logarithmic density gradient does not depend
on $x$\/ under our assumption that helium abundance is spatially uniform;
this relation will be more complicated if one uses a range of radii where
this assumption does not hold.  The electron temperature $T_e$, derived from
the shape of the X-ray spectrum, is practically independent of the helium
abundance.

%%%%%%%%%%%%%%%%%%%%%%%%%%%%%%%%%%%%%%%%%%%%%%
\begin{figure}[b]
\vspace*{5mm}
\center
\includegraphics[width=0.45\textwidth, bb=31 174 536 645]%
{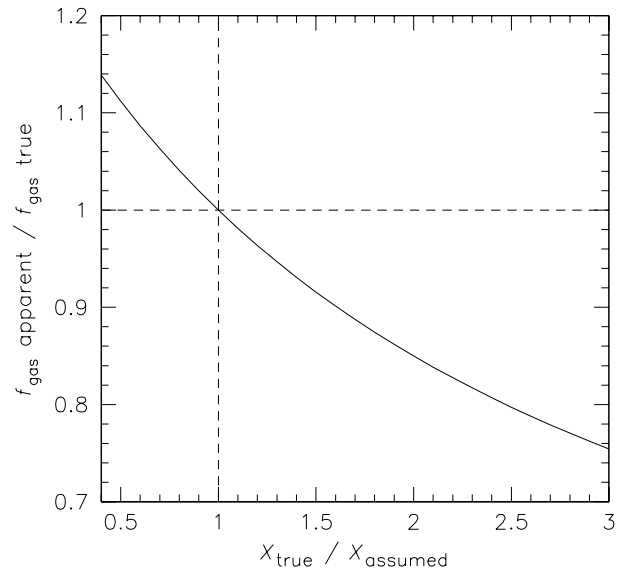}

\caption{The relative error in the X-ray derived cluster gas mass fraction,
  $f_{\rm gas}$, as a function of the error in the cluster helium abundance,
  $x$ (see eq.\ \ref{eq:fgas}).}

\label{fig:fgas}
\end{figure}
%%%%%%%%%%%%%%%%%%%%%%%%%%%%%%%%%%%%%%%%%%%%%%

Let us now consider the cluster gas mass. Provided the cluster absolute size
is known, it is proportional to the plasma density. In X-ray analysis, the
plasma hydrogen density $n_p$ is derived from the normalization of the
cluster continuum spectrum (assuming that emission lines from heavy elements
are detected and properly modeled).  The X-ray continuum luminosity is
\begin{equation}
L_X \propto n_e n_p \epsilon_{ep} (1+4x) 
         = n_p^2 \epsilon_{ep} (1+4x)(1+2x)
\end{equation}
where $\epsilon_{ep}$ is bremsstrahlung emissivity for a pure
electron-proton plasma, and the factor $(1+4x)$ accounts for the additional
bremsstrahlung on helium nuclei with charge 2. Again, heavier elements add
very little to the continuum emission. $L_X$ is the observable quantity that
one obtains directly from spectral fitting (and a known distance).  Fixing
it, we obtain the dependence of the derived $n_p$ on the assumed helium
abundance:
\begin{equation}
n_p \propto [(1+4x)(1+2x)]^{-1/2}.
\label{eq:np}
\end{equation}
The gas mass then depends on $x$\/ as follows:
\begin{equation}
M_{\rm gas} \propto n_p + 4n_{\rm He} = n_p (1+4x)
            \propto \left(\frac{1+4x}{1+2x}\right)^{1/2}.
\label{eq:mgas}
\end{equation}
Note that $M_{\rm gas}$ and $M_{\rm tot}$ change with $x$\/ in the opposite
directions, so their ratio, the gas mass fraction, depends on $x$\/ stronger
than either of these quantities:
\begin{equation}
f_{\rm gas}=\frac{M_{\rm gas}}{M_{\rm tot}} 
            \propto \frac{(1+4x)^{3/2}}{(2+3x)(1+2x)^{1/2}}.
\label{eq:fgas}
\end{equation}
We do not know the true value of $x$\/ and have to assume one to calculate
$f_{\rm gas}$; Fig.\ \ref{fig:fgas} shows the resulting relative error.
Note that the above $f_{\rm gas}(x)$ dependence is weaker than that derived
by Ettori \& Fabian (2006); the difference is due to our fixing of the
observable quantity $L_X$ when deriving $n_p$ in order to mimic the X-ray
data analysis.

Fig.\ \ref{fig:fgas} shows that an error by a factor of 2 in the assumed
helium abundance corresponds to a $\sim 15$\% error in $f_{\rm gas}$. This
is comparable to the expected difference between the apparent $f_{\rm gas}$
at $z=1$ for open and flat $\Omega_m=0.3$ cosmologies. Allen et al.\ (2004)
derived $f_{\rm gas}$ values within $r<r_{2500} \approx 0.25 r_{200}$ for a
sample of hot clusters, detected such a difference, and used it as evidence
for $\Omega_\Lambda>0$.  Of course, this cosmological test is based on a
comparison of $f_{\rm gas}$ at high and low redshifts, so the value of
helium abundance does not matter as long as it does not evolve between those
redshifts.  Furthermore, the sign of the error arising from wrongly assuming
a primordial helium abundance in the presence of sedimentation is such that
the derived $f_{\rm gas}$ would be closer to its ``pre-sedimentation''
value, which is what one would ideally want to use for such a test (A.
Vikhlinin, private communication).  A discussion of this error-cancellation
effect is beyond the scope of this paper (it would require a more accurate
calculation than that used for eq.\ \ref{eq:mtot}). However, it is clear
that any tests relying on an even higher accuracy of $f_{\rm gas}$ (such as
deriving the dark energy equation of state, which would need a few percent
accuracy on $f_{\rm gas}$) will require the knowledge of the intracluster
helium abundance at different redshifts.  We will return to this in
\S\ref{sec:da}.

%%%%%%%%%%%%%%%%%%%%%%%%%%%%%%%%%%%%%%%%%%%%%%
\begin{figure}
\center
\includegraphics[width=0.45\textwidth, bb=31 174 536 645]%
{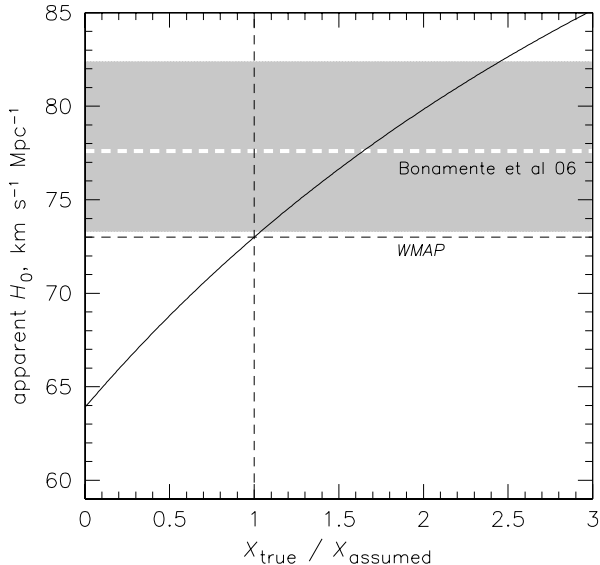}

\caption{The apparent $H_0$ value from a combination of SZ
  and X-ray cluster data, as a function of the error in the cluster helium
  abundance, $x$ (see eq.\ \ref{eq:da}), shown by solid line. For
  illustration, we take {\em WMAP}'s $H_0=73$ \kmsmpc\ as a ``true'' value
  (horizontal black dashed line; Spergel et al.\ 2007).  White dashed line
  and gray error band (statistical only, 68\%) overlays the $H_0$
  measurement by Bonamente et al.\ (2006), who assumed a helium abundance
  from Anders \& Grevesse (1989).}

\label{fig:h0}
\vspace{5mm}
\end{figure}
%%%%%%%%%%%%%%%%%%%%%%%%%%%%%%%%%%%%%%%%%%%%%%

\subsection{Hubble constant}
\label{sec:h0}

It has long been suggested (Cavaliere, Danese, \& de Zotti 1979; Silk \&
White 1978) that combining the X-ray and Sunyaev-Zeldovich (SZ) observations
of a cluster can be used to measure the absolute distance to the cluster.
This method uses the fact that the SZ decrement and the X-ray brightness
depend on different powers of the intracluster electron density, whose value
can thus be determined and converted to the distance. Below I show how this
measurement depends on the assumed helium abundance. The underlying reason
for this dependence is that the SZ effect is caused only by electrons, while
the X-ray emission is caused by scattering of electrons on protons and
helium nuclei. The fact that helium sedimentation would affect the cluster
SZ decrement was mentioned by Gilfanov \& Sunyaev (1984) but, to our
knowledge, never considered in any SZ data analyses.

In a nonrelativistic approximation, the SZ signal in the direction of a
cluster is proportional to the comptonization parameter
\begin{equation}
y \equiv \int \frac{kT_e}{m_e c^2}\,\sigma_T\, n_e(l)\, dl \,\propto\, T_e\,
n_{e0}\, d_a 
\end{equation}
(Sunyaev \& Zeldovich 1972), where the integral is along the line of sight,
$n_{e0}$ is some characteristic electron density (e.g., near the cluster
center), and $d_a$ is the angular distance to the cluster.  Here for the
rightmost part of the equation, the plasma cloud is assumed isothermal and
spherically symmetric.  The latter assumption is important (the
line-of-sight distribution of the gas density is taken to be the same as
that in the plane of the sky) but cannot be tested directly, so in practice,
a cluster sample has to be used to average out any possible ellipticities.
The surface brightness of the X-ray continuum emission from the same line of
sight is given by
\begin{equation}
S_X = \int n_e(l)\, n_p(l)\, \epsilon_{ep}\, (1+4x)\, dl \,\propto\,
     n_{e0}^2\, d_a\, \frac{1+4x}{1+2x}.
\end{equation}
The distance can be determined by combining the above equations as follows:
\begin{equation}
d_a \propto \frac{y^2}{S_X\,T_e^2}\, \frac{1+4x}{1+2x}.
\label{eq:da}
\end{equation}
The quantities $y$, $S_X$ and $T_e$ are directly measured, but helium
abundance $x$\/ has to be assumed. Fig.\ \ref{fig:h0} shows the effect of
this assumption on the derived $d_a$ (or $H_0\propto d_a^{-1}$), based on
eq.\ (\ref{eq:da}). For example, a factor of 2 error in helium abundance
results in a 10\% distance error.

\section{Measuring helium abundance in ICM}

The above dependence of the SZ--X-ray distances on the helium abundance can
be turned around and used to {\em derive}\/ the cluster helium abundance ---
provided the distance scale is known independently, for example, from
Cepheids and supernovae (Freedman et al.\ 2001; Sandage et al.\ 2006) or
from the CMB fluctuations (Spergel et al.\ 2007).  At present, the latter
two methods yield similar or smaller uncertainties on distances than those
from state of the art SZ--X-ray studies (e.g., Bonamente et al.\ 2006).
This is because measurement errors on the SZ signal are still quite large,
while this observable ($y$) enters squared in eq.\ (\ref{eq:da}).  Neither
the supernovae distances nor those from CMB fluctuations have any
significant dependence on primordial helium abundance (Ichikawa \& Takahashi
2006).

Fig.\ \ref{fig:h0} shows the value of $H_0=77.6^{+4.8}_{-4.3}$ \kmsmpc\ from
Bonamente et al.\ (2006), derived from the SZ and X-ray data on 38 clusters
at different redshifts. Of their 3 reported values, I chose the least
model-dependent one, without the hydrostatic equilibrium assumption and
excluding the central cool regions. The above error bars are 68\%
statistical-only, to illustrate an accuracy not quite achieved yet but
within immediate reach (once the instruments have been better calibrated);
their current systematic uncertainties are twice as big.

According to M. Bonamente (private communication), they assumed the Anders
\& Grevesse (1989) helium abundance, which is a factor of 1.17 higher than
the primordial value (\S\ref{sec:intro}). Had they used the primordial
abundance, their result would be $H_0\simeq 79$ \kmsmpc.  If, for the sake
of argument, we take the \wmap\ value of $H_0$ as ``true'', and attribute
the difference between these values to a helium abundance error (Fig.\ 
\ref{fig:h0}), we conclude that it should be a factor $1.9\pm0.8$ higher
than the primordial value. Of course, at this accuracy, it is consistent
with the primordial value, but we can already exclude some of the more
extreme predictions for helium sedimentation.

This and many other current SZ experiments use interferometric mapping of
the radio brightness. For typical clusters, this experimental design
effectively ``subtracts'' the signal from the shell outside $r\sim 1$ Mpc,
so the above constraint corresponds to this approximate central region.
This is about the same radius as used for the $f_{\rm gas}$ test by Allen et
al.\ (2004).

\section{SUMMARY AND DISCUSSION}
\label{sec:disc}

The usual assumption of a primordial helium abundance for the ICM is not
necessarily correct, especially in the cluster central regions where helium
may concentrate via several mechanisms, such as sedimentation. We cannot
measure the cluster helium abundance spectroscopically. As pointed out by
many authors, an incorrect helium abundance assumption will result in
incorrect cluster gas masses, total masses and gas mass fractions derived
from the X-ray data.

In this paper, we point out that cluster distances derived using the
SZ--X-ray combination also depend on helium abundance. If one gives up on
the original purpose of this method and combines it with an independent
distance scale estimator (such as supernovae or CMB fluctuations), one can
take advantage of this dependence and derive the cluster helium abundance.
At present, this seems to be the only practical way of measuring it. The
accuracy may already be interesting --- if one compares the best SZ--X-ray
value for $H_0$ (Bonamente et al.\ 2006) with the CMB value (Spergel et al.\ 
2007), the difference corresponds to a helium abundance $1.9\pm0.8$ times
the primordial value within the cluster central 1 Mpc regions (68\%
statistical-only uncertainty).  Increasing the number of clusters in the
sample and improving the SZ data accuracy will reduce this uncertainty.

Chuzhoy \& Loeb (2004) proposed another way of constraining the cluster
helium abundance --- by comparing the cluster total masses derived using the
hydrostatic equilibrium assumption (eq.\ \ref{eq:mtot} above) and any other
technique independent of helium abundance, such as gravitational lensing.
However, this approach appears less practical at present, because the
expected apparent mass difference ($\sim 10$\% for a factor of 2 error in
helium abundance) is much smaller than persistent discrepancies between
these mass estimators that are likely to be caused by substructure in the
dark matter distribution and deviations from hydrostatic equilibrium in the
ICM (e.g., Gavazzi 2005; Meneghetti et al.\ 2007). In comparison, the
SZ--X-ray method uses the same object --- the ICM --- at both wavelengths
and does not rely on hydrostatic equilibrium. It does, however, assume that
the ICM is not clumpy (which appears to be supported by \chandra\ imaging),
and requires accurate mapping of the ICM temperature structure. It also
needs spherical symmetry, which has always been an issue for the SZ method
of $H_0$ determination. It can be overcome by proper (i.e., X-ray) selection
of a sample of relaxed clusters that is also big enough to average out the
asymmetries.

Chuzhoy \& Loeb (2004) also pointed out that a higher helium abundance in
the ICM would affect stars that forms out of this ICM. Thus, evidence of
helium sedimentation may also be found in the spectra of the central cluster
galaxies.

\subsection{The $d_a(z)$ cosmological test and helium abundance}
\label{sec:da}

The main reason why we want to know the cluster helium abundances is to
remove the related uncertainty from the cluster-based cosmological tests,
such as the growth of structure tests that use cluster mass functions and
the $d_a(z)$ tests that use $f_{\rm gas}$ or the SZ--X-ray distances.
Obviously, using a competing distance estimator to measure helium abundances
will introduce degeneracies into the resulting cosmological constraints, the
detailed analysis of which is beyond the scope of this paper. In principle,
the above two distance tests can be combined to solve for the dependence on
helium abundance, because, as seen from eqs.\ (\ref{eq:fgas}) and
(\ref{eq:da}), the distances derived from $f_{\rm gas}$ ($d_a\propto f_{\rm
  gas}^{2/3}$) and from the SZ--X-ray method depend on $x$ differently. One
complication is that if one allows for helium sedimentation, the basic
assumption of the $f_{\rm gas}$-based test, that $f_{\rm gas}$ within a
certain central region does not evolve with $z$, may be violated.  So in
practice, one would have to fit together the X-ray and SZ data for a sample
of relaxed clusters spanning a range of $z$, parameterize and fit any
systematic change of helium abundance with redshift (hopefully small or
negligible), and use modeling to deduce ``pre-sedimentation'' $f_{\rm gas}$
values (to within a scaling factor), which would be the ones proportional to
the universal baryon fraction. To derive cosmological parameters other than
$H_0$, the absolute values of $d_a$ (and so the absolute values of $x$) are
not needed, only its change with $z$.  However, independent distances for at
least a few clusters will be required to determine if any helium
sedimentation occurs at all, and if so, to model it.

\acknowledgements

I thank Alexey Vikhlinin, Daisuke Nagai, Avi Loeb, and Jeremy Drake for
useful discussions.  Support for this work was provided by NASA contract
NAS8-39073 and \chandra\ grant GO6-7126X.

\end{document}